\documentclass[aps,prl,superscriptaddress,reprint,amsmath,amssymb]{revtex4-2}
\usepackage{graphicx}
\usepackage{dcolumn}
\usepackage{bm}
\usepackage{physics}
\usepackage{amsmath}
\usepackage{amssymb}
\usepackage{natbib}
\usepackage{lipsum}
\usepackage{wrapfig}
\usepackage{booktabs}
\usepackage{array}
\usepackage{float}
\usepackage{multirow}
\usepackage{braket}
\usepackage[colorlinks,linkcolor=blue,urlcolor=blue,anchorcolor=blue,citecolor=blue]{hyperref}
\usepackage{placeins}
\usepackage{cancel}

\begin{document}
\newcommand{\ud}{$\mathrm{d}$}
	
	\title{Direct observation of anisotropic exciton dispersion in the 2D semiconductor CrSBr}
	
	
    \author{Yiwen Song}
    \email{These authors contributed equally to this work.}
    \affiliation{State Key Laboratory for Artificial Microstructure $\rm{\&}$ Mesoscopic Physics and Frontiers Science Center for Nano-Optoelectronics, School of Physics, Peking University, Beijing 100871, China}
    \affiliation{Academy for Advanced Interdisciplinary Studies, Peking University, Beijing 100871, China}
    
    \author{Peiyi He}
    \email{These authors contributed equally to this work.}
    \affiliation{Electron Microscopy Laboratory, School of Physics, Peking University, Beijing 100871, China}
    \affiliation{International Center for Quantum Materials, School of Physics, Peking University, Beijing 100871, China}

    \author{Weizhe Zhang}
    \affiliation{State Key Laboratory for Artificial Microstructure $\rm{\&}$ Mesoscopic Physics and Frontiers Science Center for Nano-Optoelectronics, School of Physics, Peking University, Beijing 100871, China}

    \author{Wenyuan Ouyang}
    \affiliation{State Key Laboratory for Artificial Microstructure $\rm{\&}$ Mesoscopic Physics and Frontiers Science Center for Nano-Optoelectronics, School of Physics, Peking University, Beijing 100871, China}
    
    \author{Wenjing Liu}
    \affiliation{State Key Laboratory for Artificial Microstructure $\rm{\&}$ Mesoscopic Physics and Frontiers Science Center for Nano-Optoelectronics, School of Physics, Peking University, Beijing 100871, China}
    
    \author{Jinlong Du}
    \affiliation{Electron Microscopy Laboratory, School of Physics, Peking University, Beijing 100871, China}

    \author{Zuxin Chen}
    \affiliation{School of Semiconductor Science and Technology, South China Normal University, Foshan 528225, China}
   
    \author{Jiuyu Sun}
    \email{Contact author: sunjiuyu@njust.edu.cn}
    \affiliation{Department of Applied Physics and MIIT Key Laboratory of Semiconductor Microstructure and Quantum Sensing, Nanjing University of Science and Technology, Nanjing 210094, China}
    
    \author{Peng Gao}
    \email{Contact author: pgao@pku.edu.cn}
    \affiliation{Electron Microscopy Laboratory, School of Physics, Peking University, Beijing 100871, China}
    \affiliation{International Center for Quantum Materials, School of Physics, Peking University, Beijing 100871, China}
    \affiliation{Tsientang Institute for Advanced Study, Zhejiang 310024, China}
     
    \author{Yu Ye}
    \email{Contact author: ye$_$yu@pku.edu.cn}
    \affiliation{State Key Laboratory for Artificial Microstructure $\rm{\&}$ Mesoscopic Physics and Frontiers Science Center for Nano-Optoelectronics, School of Physics, Peking University, Beijing 100871, China}
    \affiliation{Liaoning Academy of Materials, Shenyang, 110167, China}
    \affiliation{Collaborative Innovation Center of Quantum Matter, Beijing 100871, China}
    
	%\thanks{}
	
	
	%\date{\today}
 
\begin{abstract}

We report momentum-resolved measurements of exciton dispersion in multilayer CrSBr using defocus-engineered electron energy-loss spectroscopy, supported by first-principles calculations. A pronounced in-plane anisotropy is observed, with the exciton exhibiting a linear dispersion along $\Gamma$Y within $\lvert \boldsymbol{q} \rvert < 0.007~\mathrm{\AA^{-1}}$, while remaining nearly dispersionless along $\Gamma$X. The slope reaches $7.02~\mathrm{eV\,\AA}$, among the largest reported in low-dimensional systems. The calculations reproduce the experimentally observed linear dispersion, confirming its intrinsic origin. We attribute the anisotropic dispersion to the long-range electron--hole exchange interaction, enhanced by strong out-of-plane confinement and governed by the directional selection rules of the transition dipole moment. Comparative measurements across the magnetic phase transition from the paramagnetic to the A-type antiferromagnetic state show that the dispersion remains essentially unchanged, indicating negligible coupling between exciton propagation and magnetic order. These results establish CrSBr as a model system for investigating anisotropic exciton dynamics in low-symmetry layered semiconductors.

\end{abstract}

\maketitle

% \section{\label{sec:level1}Introduction}

\begin{figure}
	\centering
	\includegraphics[width=1\linewidth]{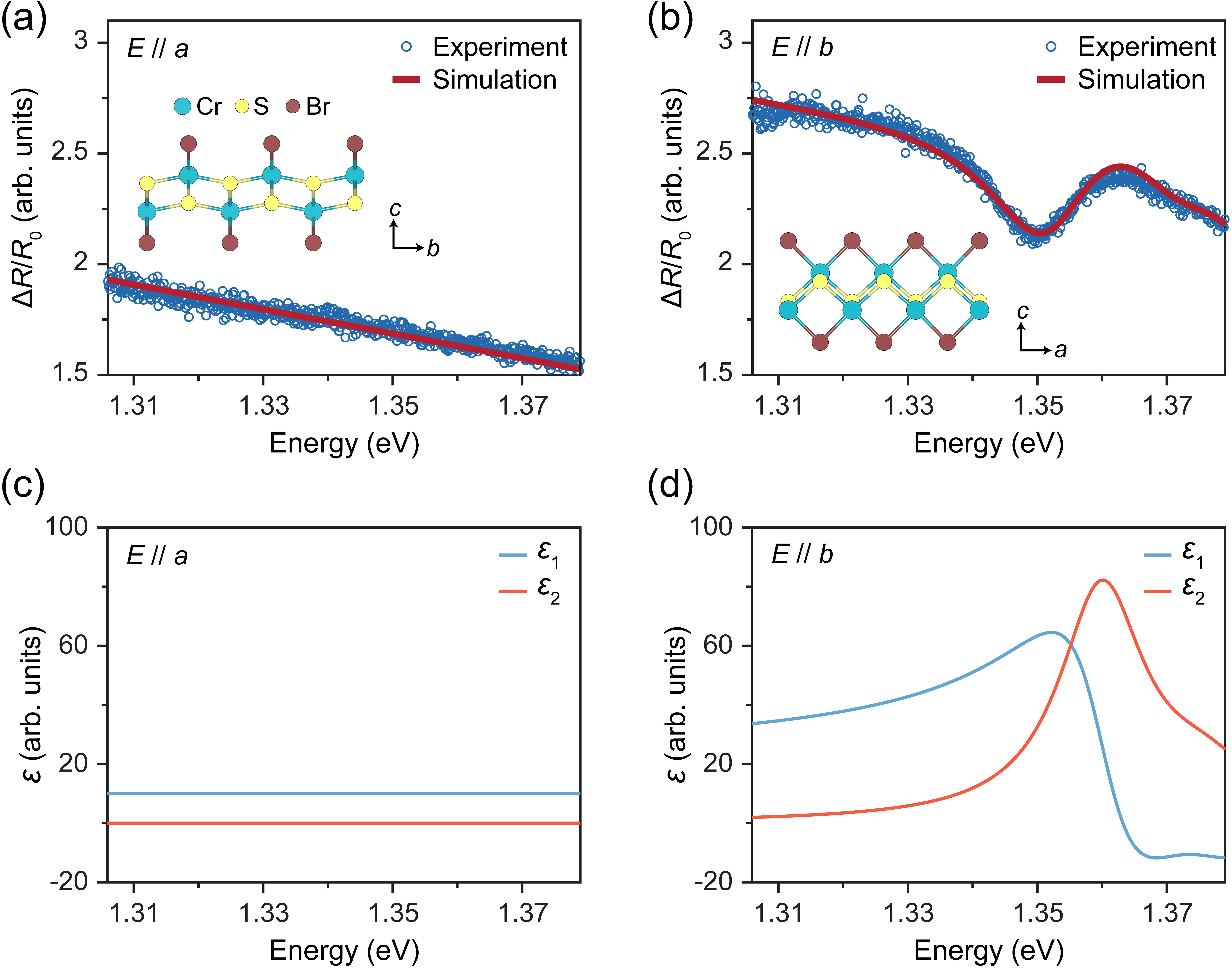}
	\caption{Polarization-resolved optical response and extracted dielectric function of CrSBr. (a, b) Experimental differential reflectance spectra (symbols) and corresponding transfer-matrix method fits (solid lines) of a 41-nm-thick CrSBr flake at $T = 100~\mathrm{K}$ under normal incidence, with light polarized along the (a) $a$ axis and (b) $b$ axis. The reference reflectance $R_0$ was obtained from a bare silicon substrate. Inset: Crystal structure of CrSBr. (c, d) Real ($\varepsilon_1$) and imaginary ($\varepsilon_2$) parts of the calculated dielectric function along the (c) $a$ axis and (d) $b$ axis.}
	\label{Figure1}
\end{figure}

Excitons, Coulomb-bound electron–hole pairs, are fundamental quasiparticles governing the optical response of semiconductors\cite{frenkel1931transformation,wannier1937structure,elliott1957intensity}. Their dispersion relation, describing the dependence of the exciton energy on the center-of-mass momentum, plays a central role in determining the exciton effective mass\cite{habenicht2015investigation,ahmadi2025exciton}, transport behavior\cite{crochet2012disorder,ghazaryan2018anisotropic}, and radiative dynamics\cite{palummo2015exciton,manolatou2016radiative,chen2019ab}. More broadly, exciton dispersion encodes essential information about the intrinsic electronic structure, dimensionality\cite{wu2015exciton,su2026dimensionality}, and symmetry\cite{cudazzo2015exciton} of a material, as well as the strength of its many-body interactions.

In conventional three-dimensional semiconductors, excitons near the zero-momentum limit typically exhibit a parabolic dispersion\cite{altarelli1977exciton,gatti2013exciton,schuster2016nongeneric}. In contrast, theoretical studies have predicted that reducing dimensionality to the two-dimensional (2D) limit can fundamentally alter excitonic behavior. Specifically, the long-range electron–hole exchange interaction introduces a nonanalytic dependence on momentum, which can give rise to a linear exciton dispersion in the small-momentum regime near the Brillouin zone center ($\Gamma$ point)\cite{qiu2015nonanalyticity,qiu2021signatures}, particularly when out-of-plane confinement enhances exchange effects. This phenomenon has been extensively discussed in the context of 2D systems such as monolayer transition metal dichalcogenides (TMDCs)\cite{wu2015exciton,qiu2015nonanalyticity} and black phosphorus\cite{cudazzo2016exciton}, where reduced screening and enhanced Coulomb interactions amplify exchange effects.

Alongside rapid progress in the synthesis and characterization of 2D materials, the experimental determination of exciton dispersion has emerged as a central challenge. Momentum-resolved electron energy-loss spectroscopy ($q$-EELS) provides a powerful approach for directly probing these excitations in energy–momentum space\cite{schuster2015anisotropic,hong2020probing,ahmadi2025exciton,su2026dimensionality,liu2026direct}. Early experimental efforts, however, were limited by insufficient energy and momentum resolution, as well as by the intrinsically narrow linear-dispersion regime in many prototypical 2D materials such as TMDCs. For instance, measurements on monolayer WSe$_2$ reported predominantly isotropic parabolic dispersions\cite{hong2020probing}, seemingly contradicting theoretical predictions. More recently, improvements in instrumental resolution, together with the identification of materials in which the exchange-induced linear dispersion becomes experimentally accessible, have enabled the direct observation of linear exciton dispersion in systems such as the quasi-2D Mott insulator Nb$_3$Cl$_8$\cite{su2026dimensionality} and monolayer \textit{h}BN\cite{liu2026direct}. These developments highlight the evolving capability of momentum-resolved spectroscopies to resolve excitonic dynamics in low-dimensional systems.

Despite these advances, most experimental studies to date have focused on systems with in-plane isotropy, leaving the role of intrinsic lattice anisotropy in shaping exciton dispersion largely unexplored. Layered CrSBr, a van der Waals magnetic semiconductor with pronounced in-plane structural anisotropy\cite{telford2020layered}, provides an ideal platform to address this question. Its highly anisotropic crystal structure\cite{ziebel2024crsbr} leads to strongly direction-dependent electronic\cite{wu2022quasi,klein2023bulk} and optical\cite{wilson2021interlayer,lee2021magnetic,sears2025observation} properties. Moreover, CrSBr undergoes a magnetic phase transition from a paramagnetic state at room temperature to an A-type antiferromagnetic state below the Néel temperature ($T_\mathrm{N} \approx 132~\mathrm{K}$)\cite{goser1990magnetic,telford2020layered,liu2022three}, offering a unique opportunity to investigate the interplay between magnetic ordering and exciton dynamics within a single material system.

In this work, we employ $q$-EELS in a scanning transmission electron microscope (STEM), complemented by first-principles many-body perturbation calculations, to investigate exciton dispersion in free-standing multilayer CrSBr across different magnetic phases. We observe a pronounced in-plane anisotropy, characterized by a linear exciton dispersion along the $\Gamma$Y direction in the small-momentum regime near the Brillouin zone center, while the $\Gamma$X direction remains nearly dispersionless. The finite-momentum Bethe--Salpeter equation (BSE) calculations faithfully reproduce the experimentally observed dispersion, confirming that it is an intrinsic property of CrSBr. Combined with an analytical model, the experiment–theory agreement identifies the linear dispersion as a consequence of long-range electron--hole exchange interaction enhanced by strong out-of-plane confinement and governed by the directional selection rules of the transition dipole moment. Furthermore, the close similarity between exciton dispersions measured in the paramagnetic and A-type antiferromagnetic phases indicates that momentum-dependent exciton dynamics are primarily determined by the intrinsic structural and electronic anisotropy, with magnetic ordering playing a negligible role. Our combined experimental and theoretical results establish the microscopic origin of the anisotropic exciton dispersion in CrSBr, providing a general framework for understanding exciton dynamics in low-symmetry layered semiconductors and for engineering directional exciton transport and anisotropy-enabled optoelectronic functionalities.

% \section{\label{sec:level2}Results and Discussions}

In transmission EELS, the scattering cross section is proportional to the loss function, $\mathrm{Im}[-1/\varepsilon(\omega, \boldsymbol{q})]$, where $\varepsilon(\omega, \boldsymbol{q})$ denotes the energy- and momentum-dependent dielectric function\cite{fink1989recent,sturm1993dynamic}. A quantitative understanding of the dielectric response of CrSBr is therefore essential for interpreting its excitonic excitations. To this end, we performed polarization-resolved differential reflectance measurements on a 41-nm-thick CrSBr flake at $T = 100~\mathrm{K}$. As shown in Fig. \ref{Figure1}(a), the reflectance spectrum for light polarized along the $a$ axis is nearly featureless in the energy range of 1.31–1.38 eV. In contrast, the $b$-axis spectrum exhibits pronounced excitonic resonances with clear absorption features [Fig. \ref{Figure1}(b)].

To quantitatively extract the dielectric function, we fitted the reflectance spectra using the transfer matrix method (TMM)\cite{Yeh1988}, modeling the complex dielectric function $\varepsilon(\omega)$ within a Lorentz oscillator formalism\cite{fujiwara2007spectroscopic}:
\begin{equation}
\varepsilon(\omega) = \varepsilon_{\mathrm{bg}} + \sum_{j} \frac{f_j}{\omega_j^2 - \omega^2 - i\Gamma_j \omega}.
\end{equation}
Here, $\varepsilon_{\mathrm{bg}}$ is the background dielectric constant, while $\omega_j$, $\Gamma_j$, and $f_j$ denote the resonance energy, linewidth, and oscillator strength of the $j$th excitation, respectively. For polarization along the $a$ axis, the spectrum is well described by a constant background with $\varepsilon_{\mathrm{bg}} = 20$, indicating negligible excitonic contribution in this energy range. In contrast, the $b$-axis response requires two excitonic oscillators, labeled $1s (\mathrm{X})$ and $\mathrm{X}^*$, centered at 1.360 eV and 1.376 eV, respectively. These resonances exhibit markedly different oscillator strengths; the $\mathrm{X}$ exciton dominates the optical response with $f_{\mathrm{X}} = 1.7~\mathrm{(eV)}^2$ and $\Gamma_{\mathrm{X}} = 16~\mathrm{meV}$, whereas the $\mathrm{X}^*$ exciton is significantly weaker, with $f_{\mathrm{X}^*} = 0.4~\mathrm{(eV)}^2$ and a linewidth of $\Gamma_{\mathrm{X}^*} = 20~\mathrm{meV}$. The resulting real ($\varepsilon_1$) and imaginary ($\varepsilon_2$) parts of the dielectric function are shown in Fig.~\ref{Figure1}(c) and Fig.~\ref{Figure1}(d). Despite the inclusion of two excitonic resonances, their small energy separation ($\sim$16~meV), comparable to their linewidths (16--20~meV), together with the strongly asymmetric oscillator strengths, renders the $\mathrm{X}^*$ resonance a lineshape modulation rather than a distinct spectral peak.

\begin{figure*}
	\centering
	\includegraphics[width=1\linewidth]{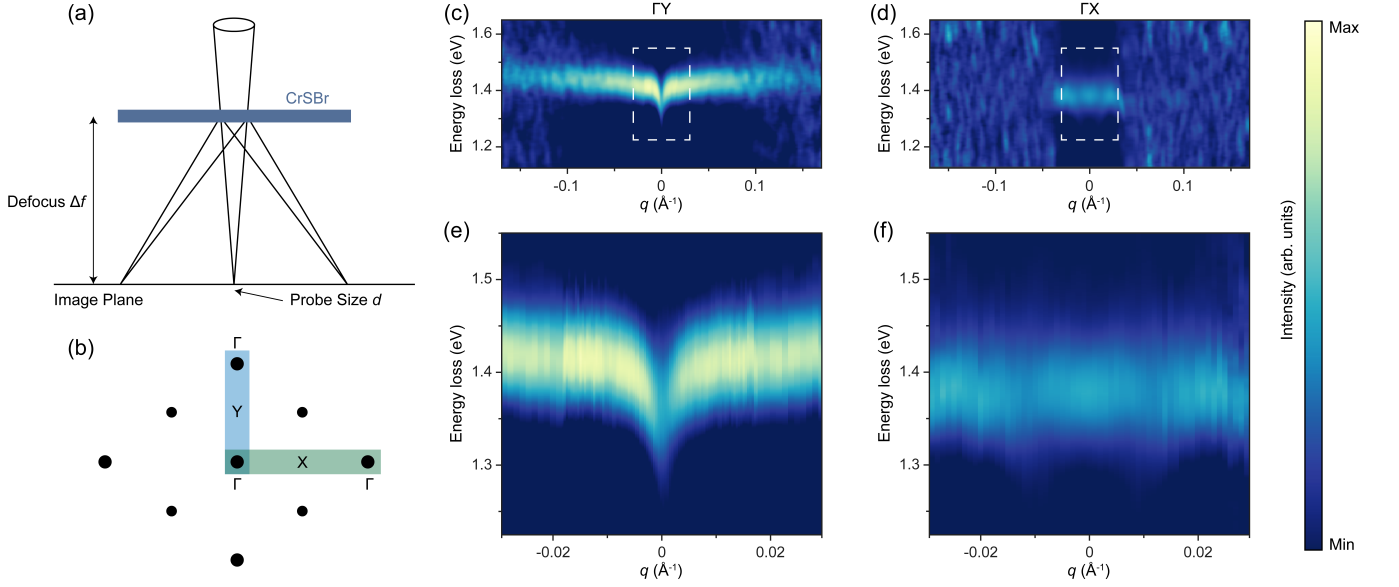}
	\caption{Defocus-engineered $q$-EELS geometry and anisotropic exciton dispersion in free-standing multilayer CrSBr at $T = 100~\mathrm{K}$. (a) Schematic illustration of the defocus-engineered $q$-EELS geometry. By displacing the sample from the focal plane, the electron beam becomes defocused at the sample position, causing electrons with different incident angles to spatially separate into discrete diffraction spots in the image plane. The momentum resolution ($\Delta q$) is tunable through the defocus distance ($\Delta f$). (b) Reciprocal-space geometry of CrSBr and the position of EELS slot aperture. The blue and green rectangles indicate the momentum paths along the $\Gamma$Y and $\Gamma$X directions, respectively. (c, d) Normalized momentum--energy maps obtained from $q$-EELS measurements in multilayer CrSBr along the (c) $\Gamma$Y and (d) $\Gamma$X directions, spanning a momentum range of $0.17~\mathrm{\AA^{-1}}$. (e, f) Magnified views of the boxed regions in (c) and (d), respectively, covering a momentum range of $0.03~\mathrm{\AA^{-1}}$ near the Brillouin-zone center.}
    
	\label{Figure2}
\end{figure*}

A pronounced in-plane anisotropy of the excitonic response in CrSBr is evident, consistent with the zero-momentum optical response framework established in our previous work\cite{wang2023magnetically}. The transition dipole moments are predominantly aligned along the crystalline $b$ axis, with negligible spectral weight along the $a$ axis. This behavior is closely linked to the in-plane anisotropic crystal structure and the quasi-one-dimensional electronic character of CrSBr\cite{wilson2021interlayer,klein2023bulk,ziebel2024crsbr}. While these zero-momentum measurements confirm the static optical anisotropy, directly probing the momentum-dependent exciton behaviors requires advanced momentum-resolved techniques.

To investigate exciton dispersion, we performed $q$-EELS measurements in a STEM operated at 30~kV with an energy resolution of approximately 18~meV. Multilayer CrSBr flakes (6--15~nm thick), prepared by mechanical exfoliation, were transferred onto holey carbon grids and measured at $T = 100~\mathrm{K}$, well below the Néel temperature, corresponding to the A-type antiferromagnetic phase. To access the small-momentum regime with sufficiently high momentum resolution, we employed a defocus-engineered $q$-EELS geometry adapted from Midgley’s approach and recently developed for low-energy dispersion measurements~\cite{do2025slow,elgvin2026advances}. In this configuration, the sample is displaced from the focal plane, causing the electron beam to become defocused at the sample position. As a result, electrons with different incident angles are spatially separated into discrete diffraction spots in the image plane [Fig.~\ref{Figure2}(a)]. Within the framework of Midgley theory~\cite{midgley1999simple}, the momentum resolution $\Delta q$ is related to the defocus distance ($\Delta f$) through $\Delta q = d / (\Delta f \lambda)$, where $d$ is the probe size and $\lambda$ is the electron wavelength. In the present experimental configuration, a convergence semi-angle of $\alpha = 8~\mathrm{mrad}$ together with a typical defocus value of $\Delta f = 200~\mathrm{\mu m}$ yields a momentum resolution of approximately $4 \times 10^{-4}~\AA^{-1}$.

Figure~\ref{Figure2}(b) illustrates the reciprocal-space configuration of CrSBr, where the orthorhombic lattice defines two principal high-symmetry directions: $\Gamma$X (along the $a$ axis) and $\Gamma$Y (along the $b$ axis). By adjusting the projector lens, the EELS slot aperture was aligned to selectively collect inelastically scattered electrons along these momentum directions, enabling direct mapping of exciton dispersion in momentum space.

To enhance weak signals at large momentum transfer and enable consistent comparison across different $\boldsymbol{q}$ values, the spectral intensity at each $\boldsymbol{q}$ was normalized over the energy range of 1.25--1.55~eV\cite{liu2023tunable}. Momentum--energy mappings along the $\Gamma$Y [Fig.~\ref{Figure2}(c)] and $\Gamma$X [Fig.~\ref{Figure2}(d)] directions exhibit mirror symmetry about the $\Gamma$ point, consistent with Brillouin zone symmetry. Along the $\Gamma$Y direction, excitonic features remain visible up to $\lvert \boldsymbol{q} \rvert \approx 0.15~\mathrm{\AA^{-1}}$ and display a pronounced blueshift with increasing momentum. In the small-momentum regime near the $\Gamma$ point, the exciton energy rises rapidly, forming a characteristic V-shaped dispersion. This behavior is further resolved in the magnified view of the $\lvert \boldsymbol{q} \rvert < 0.03~\mathrm{\AA^{-1}}$ region [Fig.~\ref{Figure2}(e)], where the steep energy variation within the central $\sim 0.01~\mathrm{\AA^{-1}}$ strongly indicates a nonparabolic dispersion, suggestive of a linear component in the small-momentum regime. In contrast, the dispersion along the $\Gamma$X direction exhibits markedly different behavior. Excitonic signals are only observable within a limited range of $\lvert \boldsymbol{q} \rvert \approx 0.04~\mathrm{\AA^{-1}}$, beyond which they become indistinguishable from the background. The corresponding magnified map [Fig.~\ref{Figure2}(f)] reveals an almost flat dispersion with negligible energy shift.

Overall, the exciton dispersion exhibits pronounced in-plane anisotropy in both the accessible momentum range and the dispersion behavior, characterized by a strong momentum-dependent blueshift along the $\Gamma$Y direction, particularly in the small-momentum regime near the $\Gamma$ point, suggesting a nonparabolic dispersion, in contrast to an essentially dispersionless response along the $\Gamma$X direction.

\begin{figure}
	\centering
	\includegraphics[width=1\linewidth]{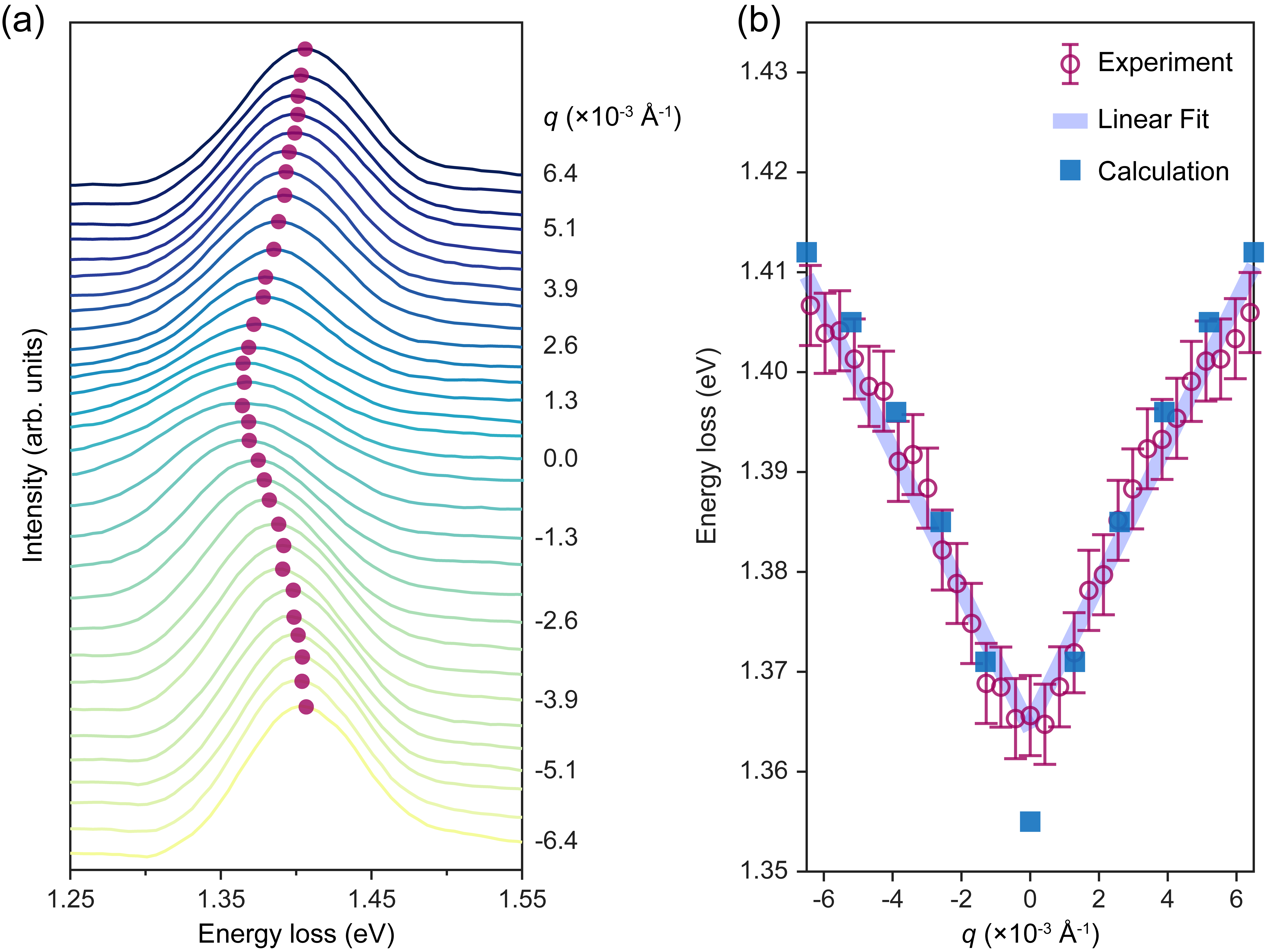}
	\caption{Linear exciton dispersion in multilayer CrSBr near the $\Gamma$ point along the $\Gamma$Y direction at $T = 100~\mathrm{K}$. (a) Momentum-resolved EELS spectra showing exciton peak evolution along $\Gamma$Y. Magenta symbols denote peak positions from Gaussian fitting; corresponding momentum values ($q$) are on the right axis. (b) Exciton energy as a function of momentum $q$. Magenta circles represent the experimental peak positions extracted from $q$-EELS measurements, while blue squares denote the finite-momentum Bethe--Salpeter equation (BSE) calculations. The purple line is a linear fit to the experimental data, demonstrating the linear dispersion in the small-momentum regime.}
	\label{Figure3}
\end{figure}

To quantitatively characterize the exciton dispersion, we extracted momentum-resolved spectra from the $q$-EELS mappings within a momentum range of $\pm 0.1~\mathrm{\AA^{-1}}$ around the $\Gamma$ point along both the $\Gamma$X and $\Gamma$Y directions. Exciton peak positions were determined by Gaussian fitting [Supplemental Material, Fig.~S1], further confirming the pronounced anisotropy of the dispersion. Figure~\ref{Figure3}(a) shows representative spectra along the $\Gamma$Y direction in the small-momentum regime, with magenta symbols marking the extracted peak positions. The corresponding exciton energy as a function of $q$ is plotted in Fig.~\ref{Figure3}(b), where a linear fit (purple line) captures the dispersion near the $\Gamma$ point. A clear linear dependence is observed in the small-momentum regime, followed by a gradual deviation at larger $\lvert \boldsymbol{q} \rvert$ [Supplemental Material, Fig.~S2]. In addition to the experimental data, the finite-momentum Bethe–Salpeter equation (BSE) calculations are overlaid for comparison [computational details described in Supplemental Material]. The calculated exciton dispersion reproduces the linear energy–momentum dependence in the small-momentum regime along the $\Gamma$Y direction. This excellent agreement confirms that the observed dispersion is an intrinsic property of CrSBr and provides a solid basis for understanding its microscopic origin.

The linear dispersion along the $\Gamma$Y direction originates from the interplay between the orientational selection rules of the transition dipole moment and the long-range electron--hole exchange interaction. As established in Ref.~\cite{qiu2021signatures}, such exchange interaction in 2D systems introduces a nonanalytic linear term in the exciton dispersion, which for nondegenerate excitons can be generally expressed as
\begin{equation}
E(\boldsymbol{q}) = E_0 + A \lvert \boldsymbol{q} \rvert \cos^2 \theta_{\boldsymbol{q}} + \frac{\hbar^2 q_x^2}{2M_x^*} + \frac{\hbar^2 q_y^2}{2M_y^*},
\end{equation}
where $E_0$ is the exciton energy at the $\Gamma$ point, $A$ characterizes the strength of the long-range exchange interaction, $\boldsymbol{q}$ is the center-of-mass momentum, and $\theta_{\boldsymbol{q}}$ denotes the angle between $\boldsymbol{q}$ and the transition dipole moment. Here, $M_x^*$ and $M_y^*$ represent the effective masses of the exciton center-of-mass motion along the $x$ and $y$ directions, respectively.

In CrSBr, the transition dipole moment is aligned along the crystalline $b$ axis ($y$ direction), yielding $\cos^2 \theta_{\boldsymbol{q}} = q_y^2 / \lvert \boldsymbol{q} \rvert^2$. Substituting into Eq.~(2) gives
\begin{equation}
E(\boldsymbol{q}) = E_0 + A \frac{q_y^2}{\lvert \boldsymbol{q} \rvert} + \frac{\hbar^2 q_x^2}{2M_x^*} + \frac{\hbar^2 q_y^2}{2M_y^*}.
\end{equation}
Along the $\Gamma$Y direction ($q_x = 0$), this expression reduces to
\begin{equation}
E(0, q_y) = E_0 + A \lvert q_y \rvert + \frac{\hbar^2 q_y^2}{2M_y^*}.
\end{equation}
In the small-$q_y$ limit, the linear term dominates, giving rise to the observed V-shaped dispersion. The observed linear dispersion reflects the dominance of long-range exchange interaction, enabled by strong out-of-plane confinement, rather than being a direct consequence of reduced dimensionality. Previous first-principles calculations reveal that the exciton wavefunction is strongly confined within individual CrSBr layers, with weak interlayer coupling\cite{wilson2021interlayer}. Such out-of-plane localization enhances the long-range exchange interaction in CrSBr, leading to the observed small-momentum linear dispersion even in this multilayer system.

Furthermore, the slope of the linear dispersion corresponds to $\hbar$ times the exciton group velocity, with $v = (1/\hbar)\, dE/dq$\cite{toyozawa1959dynamical,thompson2022anisotropic}, providing a direct measure of the strength of the long-range exchange interaction. The extracted slope along the $\Gamma$Y direction is $7.02~\mathrm{eV\,\AA}$, significantly exceeding that reported for monolayer MoS$_2$ ($1.8~\mathrm{eV\,\AA}$)\cite{qiu2015nonanalyticity}, $\alpha$-Nb$_3$Cl$_8$ ($0.51~\mathrm{eV\,\AA}$)\cite{su2026dimensionality}, and monolayer \textit{h}BN ($6.08~\mathrm{eV\,\AA}$ along $\Gamma$M and $4.22~\mathrm{eV\,\AA}$ along $\Gamma$K)\cite{liu2026direct}. The unusually large exciton group velocity in CrSBr reflects an anisotropy-enhanced nonanalytic long-range exchange interaction arising from the directional alignment of transition dipole moments. Such a high group velocity establishes an intrinsic velocity scale for exciton propagation, potentially enabling efficient energy transport over mesoscopic length scales.

To further investigate the interplay between magnetic order and exciton dynamics, we performed identical $q$-EELS measurements on multilayer CrSBr in the paramagnetic state at $T = 300~\mathrm{K}$ [Supplemental Material, Figs.~S3--S5]. A direct comparison between the paramagnetic and A-type antiferromagnetic phases reveals that the exciton dispersion is remarkably robust across the magnetic transition. In particular, the anisotropic dispersion characteristics remain essentially unchanged; the small-momentum linear dispersion along the $\Gamma$Y direction persists, while the $\Gamma$X direction continues to exhibit negligible momentum dependence. Within experimental resolution, no discernible renormalization of the dispersion is observed.

A strong coupling between exciton propagation and spin degrees of freedom\cite{wang2023magnetically,smiertka2026distinct} can lead to a reconstruction of the dispersion across a magnetic phase transition, manifested as momentum-dependent renormalization of the dispersion arising from exchange-driven modifications of the electronic structure and spin-dependent band splitting\cite{Auerbach_1994}. Such signatures are absent in our measurements. The absence of discernible changes indicates that magnetic ordering has a negligible influence on the exciton dispersion, implying a weak coupling between the exciton center-of-mass motion and the underlying spin system. Instead, the dispersion is primarily governed by the intrinsic in-plane lattice and electronic anisotropy.

% \section{Conclusion}

%\section*{Conclusion and Discussion}

In summary, we combine defocus-engineered $q$-EELS with first-principles calculations to resolve the momentum-dependent exciton dispersion in multilayer CrSBr. The measurements reveal a pronounced in-plane anisotropy, characterized by a linear dispersion along the $\Gamma$Y direction in the small-momentum regime and a nearly dispersionless response along $\Gamma$X. The excellent agreement between experiment and theory identifies the observed anisotropic dispersion as an intrinsic property of CrSBr, originating from the long-range electron--hole exchange interaction enhanced by strong out-of-plane confinement and governed by the directional selection rules of the transition dipole moment. Despite the presence of magnetic ordering, the exciton dispersion remains essentially unchanged across the magnetic phase transition, indicating negligible coupling between exciton center-of-mass motion and the underlying spin system.

More broadly, our work establishes a general route for engineering exciton dynamics through the interplay of exchange interaction, confinement, and lattice anisotropy. This provides a framework for controlling exciton transport in low-symmetry layered semiconductors and offers new opportunities for anisotropy-driven optoelectronic functionalities.

\begin{acknowledgments}
This work was supported by the National Key R\&D Program of China (Nos. 2025YFA1411002 and 2025YFA1411001), the National Natural Science Foundation of China (Nos. 12425402 and 22303041), and the Fundamental Research Funds for the Central Universities (No.30923010203). The authors acknowledge support from the High Performance Computing Platform of Nanjing University of Aeronautics and Astronautics. P.G. acknowledges the support from the New Cornerstone Science Foundation through the XPLORER PRIZE.

\end{acknowledgments}

\bibliographystyle{apsrev4-2}
\bibliography{Reference} 

@article{frenkel1931transformation,
  title={On the transformation of light into heat in solids. I},
  author={Frenkel, Jacov},
  journal={Physical Review},
  volume={37},
  number={1},
  pages={17},
  year={1931},
  publisher={APS},
  URL={https://doi.org/10.1103/PhysRev.37.17},
}

@article{wannier1937structure,
  title={The structure of electronic excitation levels in insulating crystals},
  author={Wannier, Gregory H},
  journal={Physical Review},
  volume={52},
  number={3},
  pages={191},
  year={1937},
  publisher={APS},
  URL={https://doi.org/10.1103/PhysRev.52.191},
}

@article{elliott1957intensity,
  title={Intensity of optical absorption by excitons},
  author={Elliott, Roger J},
  journal={Physical Review},
  volume={108},
  number={6},
  pages={1384},
  year={1957},
  publisher={APS},
  URL={https://doi.org/10.1103/PhysRev.108.1384},
}

@article{qiu2015nonanalyticity,
  title={Nonanalyticity, valley quantum phases, and lightlike exciton dispersion in monolayer transition metal dichalcogenides: Theory and first-principles calculations},
  author={Qiu, Diana Y and Cao, Ting and Louie, Steven G},
  journal={Physical review letters},
  volume={115},
  number={17},
  pages={176801},
  year={2015},
  publisher={APS},
  URL={https://doi.org/10.1103/PhysRevLett.115.176801},
}

@article{habenicht2015investigation,
  title={Investigation of the dispersion and the effective masses of excitons in bulk 2H-MoS2 using transition electron energy-loss spectroscopy},
  author={Habenicht, Carsten and Knupfer, Martin and B{\"u}chner, Bernd},
  journal={Physical Review B},
  volume={91},
  number={24},
  pages={245203},
  year={2015},
  publisher={APS},
  URL={https://doi.org/10.1103/PhysRevB.91.245203},
}

@article{ahmadi2025exciton,
  title={Exciton dispersion and effective mass in 2H-WSe2},
  author={Ahmadi Heidari, Zahra and Knupfer, Martin},
  journal={Physical Review B},
  volume={111},
  number={20},
  pages={205205},
  year={2025},
  publisher={APS},
  URL={https://doi.org/10.1103/PhysRevB.111.205205},
}

@article{crochet2012disorder,
  title={Disorder limited exciton transport in colloidal single-wall carbon nanotubes},
  author={Crochet, Jared J and Duque, Juan G and Werner, James H and Lounis, Brahim and Cognet, Laurent and Doorn, Stephen K},
  journal={Nano letters},
  volume={12},
  number={10},
  pages={5091--5096},
  year={2012},
  publisher={ACS Publications},
  URL={https://doi.org/10.1021/nl301739d},
}

@article{ghazaryan2018anisotropic,
  title={Anisotropic exciton transport in transition-metal dichalcogenides},
  author={Ghazaryan, Areg and Hafezi, Mohammad and Ghaemi, Pouyan},
  journal={Physical Review B},
  volume={97},
  number={24},
  pages={245411},
  year={2018},
  publisher={APS},
  URL={https://doi.org/10.1103/PhysRevB.97.245411},
}

@article{qiu2021signatures,
  title={Signatures of dimensionality and symmetry in exciton band structure: Consequences for exciton dynamics and transport},
  author={Qiu, Diana Y and Cohen, Galit and Novichkova, Dana and Refaely-Abramson, Sivan},
  journal={Nano letters},
  volume={21},
  number={18},
  pages={7644--7650},
  year={2021},
  publisher={ACS Publications},
  URL={https://doi.org/10.1021/acs.nanolett.1c02352},
}

@article{palummo2015exciton,
  title={Exciton radiative lifetimes in two-dimensional transition metal dichalcogenides},
  author={Palummo, Maurizia and Bernardi, Marco and Grossman, Jeffrey C},
  journal={Nano Letters},
  volume={15},
  number={5},
  pages={2794--2800},
  year={2015},
  publisher={ACS Publications},
  URL={https://doi.org/10.1021/nl503799t},
}

@article{manolatou2016radiative,
  title={Radiative and nonradiative exciton energy transfer in monolayers of two-dimensional group-VI transition metal dichalcogenides},
  author={Manolatou, Christina and Wang, Haining and Chan, Weimin and Tiwari, Sandip and Rana, Farhan},
  journal={Physical Review B},
  volume={93},
  number={15},
  pages={155422},
  year={2016},
  publisher={APS},
  URL={https://doi.org/10.1103/PhysRevB.93.155422},
}

@article{chen2019ab,
  title={Ab initio calculations of exciton radiative lifetimes in bulk crystals, nanostructures, and molecules},
  author={Chen, Hsiao-Yi and Jhalani, Vatsal A and Palummo, Maurizia and Bernardi, Marco},
  journal={Physical Review B},
  volume={100},
  number={7},
  pages={075135},
  year={2019},
  publisher={APS},
  URL={https://doi.org/10.1103/PhysRevB.100.075135},
}

@article{cudazzo2015exciton,
  title={Exciton dispersion in molecular solids},
  author={Cudazzo, Pierluigi and Sottile, Francesco and Rubio, Angel and Gatti, Matteo},
  journal={Journal of Physics: Condensed Matter},
  volume={27},
  number={11},
  pages={113204},
  year={2015},
  publisher={IOP Publishing},
  URL={http://dx.doi.org/10.1088/0953-8984/27/11/113204},
}

@article{altarelli1977exciton,
  title={Exciton dispersion in semiconductors with degenerate bands},
  author={Altarelli, M and Lipari, NO},
  journal={Physical Review B},
  volume={15},
  number={10},
  pages={4898},
  year={1977},
  publisher={APS},
  URL={https://doi.org/10.1103/PhysRevB.15.4898},
}

@article{gatti2013exciton,
  title={Exciton dispersion from first principles},
  author={Gatti, Matteo and Sottile, Francesco},
  journal={Physical Review B—Condensed Matter and Materials Physics},
  volume={88},
  number={15},
  pages={155113},
  year={2013},
  publisher={APS},
  URL={https://doi.org/10.1103/PhysRevB.88.155113},
}

@article{schuster2016nongeneric,
  title={Nongeneric dispersion of excitons in the bulk of WSe2},
  author={Schuster, R and Wan, Y and Knupfer, M and B{\"u}chner, B},
  journal={Physical Review B},
  volume={94},
  number={8},
  pages={085201},
  year={2016},
  publisher={APS},
  URL={https://doi.org/10.1103/PhysRevB.94.085201},
}

@article{wu2015exciton,
  title={Exciton band structure of monolayer MoS2},
  author={Wu, Fengcheng and Qu, Fanyao},
  journal={Physical Review B},
  volume={91},
  number={7},
  pages={075310},
  year={2015},
  publisher={APS},
  URL={https://doi.org/10.1103/PhysRevB.91.075310},
}

@article{cudazzo2016exciton,
  title={Exciton band structure in two-dimensional materials},
  author={Cudazzo, Pierluigi and Sponza, Lorenzo and Giorgetti, Christine and Reining, Lucia and Sottile, Francesco and Gatti, Matteo},
  journal={Physical review letters},
  volume={116},
  number={6},
  pages={066803},
  year={2016},
  publisher={APS},
  URL={https://doi.org/10.1103/PhysRevLett.116.066803},
}

@article{schuster2015anisotropic,
  title={Anisotropic particle-hole excitations in black phosphorus},
  author={Schuster, R and Trinckauf, J and Habenicht, C and Knupfer, M and B{\"u}chner, B},
  journal={Physical Review Letters},
  volume={115},
  number={2},
  pages={026404},
  year={2015},
  publisher={APS},
  URL={https://doi.org/10.1103/PhysRevLett.115.026404},
}

@article{hong2020probing,
  title={Probing exciton dispersions of freestanding monolayer WSe2 by momentum-resolved electron energy-loss spectroscopy},
  author={Hong, Jinhua and Senga, Ryosuke and Pichler, Thomas and Suenaga, Kazu},
  journal={Physical review letters},
  volume={124},
  number={8},
  pages={087401},
  year={2020},
  publisher={APS},
  URL={https://doi.org/10.1103/PhysRevLett.124.087401},
}

@article{su2026dimensionality,
  title={Dimensionality-Dependent Exciton Dispersion in a Single-Band Mott Insulator},
  author={Su, Zhibin and Mi, Junjian and Yan, Shaohua and Li, Jiade and Xue, Siwei and Tao, Zhiyu and Wang, Enling and Shi, Xiongfei and Lei, Hechang and Xu, Zhuan and Guo, Jiandong and Zhu, Xuetao},
  journal={Physical Review Letters},
  volume={136},
  number={10},
  pages={106502},
  year={2026},
  publisher={APS},
  URL={https://doi.org/10.1103/1z8v-hvkf},
}

@article{liu2026direct,
  title={Direct observation of massless excitons and linear exciton dispersion},
  author={Liu, Luna Y and Woo, Steffi Y and Wu, Jinyuan and Hou, Bowen and Su, Cong and Qiu, Diana Y},
  journal={Nature Physics},
  pages={1--6},
  year={2026},
  publisher={Nature Publishing Group},
  URL={https://doi.org/10.1038/s41567-026-03193-8},
}

@article{telford2020layered,
  title={Layered antiferromagnetism induces large negative magnetoresistance in the van der Waals semiconductor CrSBr},
  author={Telford, Evan J and Dismukes, Avalon H and Lee, Kihong and Cheng, Minghao and Wieteska, Andrew and Bartholomew, Amymarie K and Chen, Yu-Sheng and Xu, Xiaodong and Pasupathy, Abhay N and Zhu, Xiaoyang and Dean, Cory R and Roy, Xavier},
  journal={Advanced Materials},
  volume={32},
  number={37},
  pages={2003240},
  year={2020},
  publisher={Wiley Online Library},
  URL={ https://doi.org/10.1002/adma.202003240},
}

@article{ziebel2024crsbr,
  title={CrSBr: an air-stable, two-dimensional magnetic semiconductor},
  author={Ziebel, Michael E and Feuer, Margalit L and Cox, Jordan and Zhu, Xiaoyang and Dean, Cory R and Roy, Xavier},
  journal={Nano Letters},
  volume={24},
  number={15},
  pages={4319--4329},
  year={2024},
  publisher={ACS Publications},
  URL={https://doi.org/10.1021/acs.nanolett.4c00624},
}

@article{wu2022quasi,
  title={Quasi-1D electronic transport in a 2D magnetic semiconductor},
  author={Wu, Fan and Guti{\'e}rrez-Lezama, Ignacio and L{\'o}pez-Paz, Sara A and Gibertini, Marco and Watanabe, Kenji and Taniguchi, Takashi and von Rohr, Fabian O and Ubrig, Nicolas and Morpurgo, Alberto F},
  journal={Advanced Materials},
  volume={34},
  number={16},
  pages={2109759},
  year={2022},
  publisher={Wiley Online Library},
  URL={ https://doi.org/10.1002/adma.202109759},
}

@article{klein2023bulk,
  title={The bulk van der Waals layered magnet CrSBr is a quasi-1D material},
  author={Klein, Julian and Pingault, Benjamin and Florian, Matthias and Hei{\ss}enb{\"u}ttel, Marie-Christin and Steinhoff, Alexander and Song, Zhigang and Torres, Kierstin and Dirnberger, Florian and Curtis, Jonathan B and Weile, Mads and Penn, Aubrey and Deilmann, Thorsten and Dana, Rami and Bushati, Rezlind and Quan, Jiamin and Luxa, Jan and Sofer, Zden\v{e}k and Al{\'u}, Andrea and Menon, Vinod M and Wurstbauer, Ursula and Rohlfing, Michael and Narang, Prineha and Lon\v{c}ar, Marko and Ross, Frances M},
  journal={ACS Nano},
  volume={17},
  number={6},
  pages={5316--5328},
  year={2023},
  publisher={ACS Publications},
  URL={https://doi.org/10.1021/acsnano.2c07316},
}

@article{wilson2021interlayer,
  title={Interlayer electronic coupling on demand in a 2D magnetic semiconductor},
  author={Wilson, Nathan P and Lee, Kihong and Cenker, John and Xie, Kaichen and Dismukes, Avalon H and Telford, Evan J and Fonseca, Jordan and Sivakumar, Shivesh and Dean, Cory and Cao, Ting and  Roy, Xavier and Xu, Xiaodong and Zhu, Xiaoyang},
  journal={Nature Materials},
  volume={20},
  number={12},
  pages={1657--1662},
  year={2021},
  publisher={Nature Publishing Group UK London},
  URL={https://doi.org/10.1038/s41563-021-01070-8},
}

@article{lee2021magnetic,
  title={Magnetic order and symmetry in the 2D semiconductor CrSBr},
  author={Lee, Kihong and Dismukes, Avalon H and Telford, Evan J and Wiscons, Ren A and Wang, Jue and Xu, Xiaodong and Nuckolls, Colin and Dean, Cory R and Roy, Xavier and Zhu, Xiaoyang},
  journal={Nano Letters},
  volume={21},
  number={8},
  pages={3511--3517},
  year={2021},
  publisher={ACS Publications},
  URL={https://doi.org/10.1021/acs.nanolett.1c00219},
}

@article{sears2025observation,
  title={Observation of Anisotropic Dispersive Dark-Exciton Dynamics in CrSBr},
  author={Sears, J and Zager, B and He, W and Occhialini, CA and Shen, Y and Lajer, M and Villanova, JW and Berlijn, T and Yakhou-Harris, F and Brookes, NB and others},
  journal={Physical Review Letters},
  volume={135},
  number={14},
  pages={146503},
  year={2025},
  publisher={APS},
  URL={https://doi.org/10.1103/fz3h-6jdx},
}

@article{goser1990magnetic,
  title={Magnetic properties of CrSBr},
  author={G{\"o}ser, O and Paul, W and Kahle, HG},
  journal={Journal of magnetism and magnetic materials},
  volume={92},
  number={1},
  pages={129--136},
  year={1990},
  publisher={Elsevier},
  URL={https://doi.org/10.1016/0304-8853(90)90689-N},
}

@article{liu2022three,
  title={A three-stage magnetic phase transition revealed in ultrahigh-quality van der Waals bulk magnet CrSBr},
  author={Liu, Wenhao and Guo, Xiaoyu and Schwartz, Jonathan and Xie, Hongchao and Dhale, Nikhil Uday and Sung, Suk Hyun and Kondusamy, Aswin Lakshmi Narayanan and Wang, Xiqu and Zhao, Haonan and Berman, Diana and Hovden, Robert and Zhao, Liuyan and Lv, Bing},
  journal={ACS Nano},
  volume={16},
  number={10},
  pages={15917--15926},
  year={2022},
  publisher={ACS Publications},
  URL={https://doi.org/10.1021/acsnano.2c02896},
}

@article{fink1989recent,
  title={Recent developments in energy-loss spectroscopy},
  author={Fink, J{\"o}rg},
  journal={Advances in electronics and electron physics},
  volume={75},
  pages={121--232},
  year={1989},
  publisher={Elsevier},
  URL={https://doi.org/10.1016/S0065-2539(08)60947-6},
}

@article{sturm1993dynamic,
  title={Dynamic structure factor: An introduction},
  author={Sturm, K},
  journal={Zeitschrift f{\"u}r Naturforschung A},
  volume={48},
  number={1-2},
  pages={233--242},
  year={1993},
  publisher={Verlag der Zeitschrift f{\"u}r Naturforschung},
  URL={https://doi.org/10.1515/zna-1993-1-244},
}

@book{Yeh1988,
  author    = {Yeh, Pochi},
  title     = {Optical Waves in Layered Media},
  publisher = {Wiley},
  address   = {New York},
  year      = {1988}
}

@book{fujiwara2007spectroscopic,
  author={Fujiwara, Hiroyuki},
  title     = {Spectroscopic Ellipsometry: Principles and Applications},
  publisher = {John Wiley \& Sons},
  address   = {Chichester},
  year      = {2007}
}

@article{do2025slow,
  title={Slow and highly confined plasmons observed in atomically thin TaS2},
  author={Do, Hue TB and Zhao, Meng and Li, Pengfei and Soh, Yu Wei and Rangaraj, Jagadesh and Liu, Bingyan and Jiang, Tianyu and Zhang, Xinyue and Lu, Jiong and Song, Peng and Teng, Jinghua and Bosman, Michel},
  journal={Nature Communications},
  volume={16},
  number={1},
  pages={5801},
  year={2025},
  publisher={Nature Publishing Group UK London},
  URL={https://doi.org/10.1038/s41467-025-60814-1},
}

@article{elgvin2026advances,
  title={Advances in momentum-resolved EELS of phonons, excitons and plasmons in 2D materials and their heterostructures},
  author={Elgvin, Cana and Hage, Fredrik S and Prytz, {\O}ystein and Elyas, Khairi and H{\"o}flich, Katja and Koch, Christoph T and Nerl, Hannah C},
  journal={Physical Review Materials},
  volume={10},
  number={2},
  pages={020201},
  year={2026},
  publisher={APS},
  URL={https://doi.org/10.1103/8wwb-lxk2},
}

@article{midgley1999simple,
  title={A simple new method to obtain high angular resolution $\omega$--q patterns},
  author={Midgley, PA},
  journal={Ultramicroscopy},
  volume={76},
  number={3},
  pages={91--96},
  year={1999},
  publisher={Elsevier},
  URL={https://doi.org/10.1016/S0304-3991(98)00088-6},
}

@article{liu2023tunable,
  title={Tunable interband transitions in twisted h-BN/graphene heterostructures},
  author={Liu, Bingyao and Zhang, Yu-Tian and Qiao, Ruixi and Shi, Ruochen and Li, Yuehui and Guo, Quanlin and Li, Jiade and Li, Xiaomei and Wang, Li and Qi, Jiajie and Du, Shixuan and Ren, Xinguo and Liu, Kaihui and Gao, Peng and Zhang, Yu-Yang},
  journal={Physical Review Letters},
  volume={131},
  number={1},
  pages={016201},
  year={2023},
  publisher={APS},
  URL={https://doi.org/10.1103/PhysRevLett.131.016201},
}

@article{toyozawa1959dynamical,
  title={On the dynamical behavior of an exciton},
  author={Toyozawa, Yutaka},
  journal={Progress of Theoretical Physics Supplement},
  volume={12},
  pages={111--140},
  year={1959},
  publisher={Oxford University Press},
  URL={https://doi.org/10.1143/PTPS.12.111},
}

@article{thompson2022anisotropic,
  title={Anisotropic exciton diffusion in atomically-thin semiconductors},
  author={Thompson, Joshua JP and Brem, Samuel and Verjans, Marne and Schmidt, Robert and Michaelis de Vasconcellos, Steffen and Bratschitsch, Rudolf and Malic, Ermin},
  journal={2D Materials},
  volume={9},
  number={2},
  pages={025008},
  year={2022},
  publisher={IOP Publishing},
  URL={https://doi.org/10.1088/2053-1583/ac4d13},
}

@article{wang2023magnetically,
  title={Magnetically-dressed CrSBr exciton-polaritons in ultrastrong coupling regime},
  author={Wang, Tingting and Zhang, Dingyang and Yang, Shiqi and Lin, Zhongchong and Chen, Quan and Yang, Jinbo and Gong, Qihuang and Chen, Zuxin and Ye, Yu and Liu, Wenjing},
  journal={Nature Communications},
  volume={14},
  number={1},
  pages={5966},
  year={2023},
  publisher={Nature Publishing Group UK London},
  URL={https://doi.org/10.1038/s41467-023-41688-7},
}

@article{smiertka2026distinct,
  title={Distinct magneto-optical response of Frenkel and Wannier excitons in CrSBr},
  author={{\'S}miertka, Maciej and Ryga{\l}a, Micha{\l} and Posmyk, Katarzyna and Peksa, Paulina and Dyksik, Mateusz and Pashov, Dimitar and Mosina, Kseniia and Sofer, Zden{\v{e}}k and van Schilfgaarde, Mark and Dirnberger, Florian and Baranowski, Micha{\l} and Acharya, Swagata and Plochocka, Paulina},
  journal={Nature Communications},
  year={2026},
  publisher={Nature Publishing Group UK London},
  URL={https://doi.org/10.1038/s41467-026-68482-5},
}

@book{Auerbach_1994,
  author    = {Assa Auerbach},
  title     = {Interacting Electrons and Quantum Magnetism},
  publisher = {Springer},
  address   = {New York},
  year      = {1994}
}
\end{document}